\documentstyle[preprint,aps,eqsecnum,epsfig]{revtex} 
\tighten 
\begin{document} 
\preprint{PITT-98-; LPTHE-98-} 
\draft  
\title{\bf ANOMALOUS KINETICS OF HARD CHARGED PARTICLES: DYNAMICAL
RENORMALIZATION GROUP RESUMMATION}   
\author{\bf D. Boyanovsky$^{(a)}$, H. J. de Vega$^{(b)}$}
\address
{(a) Department of Physics and Astronomy, University of 
Pittsburgh, Pittsburgh  PA. 15260, U.S.A \\ 
(b) LPTHE, Universit\'e Pierre et Marie Curie (Paris VI) et Denis Diderot 
(Paris VII), Tour 16, 1er. \'etage, 4, Place Jussieu, 75252 Paris, Cedex 05, 
FRANCE. Laboratoire Associ\'e au CNRS UMR 7589.}  
\date{\today}
\maketitle 
\begin{abstract} 
We study the kinetics of the distribution function for charged
particles of hard momentum in scalar QED. The goal is to understand
the effects of 
infrared divergences associated with the exchange of quasistatic
magnetic photons in the relaxation of the distribution function. We
begin by obtaining a kinetic 
transport equation for the distribution function for hard charged
scalars in a perturbative expansion that includes hard thermal loop
resummation. Solving this transport equation, the infrared divergences arising
from absorption and emission of soft quasi-static magnetic photons are
manifest in logarithmic secular terms. We
then implement the dynamical renormalization group resummation of
these secular terms in the relaxation time approximation. The
distribution function (in the linearized regime) is found to approach
equilibrium as 
$\delta n_k(t) =\delta n_k(t_o)\;  e^{-2\alpha T (t-t_o)
\ln[(t-t_o)\bar{\mu}]}$,  
with $\bar{\mu}\approx \omega_p$ the plasma frequency and $ \alpha =
e^2/4\pi $. This anomalous
relaxation is recognized to be the {\em square} of the relaxation of the
single particle propagator, providing a generalization of the usual
relation between the damping and the interaction rate. The
renormalization group approach to kinetics reveals clearly the time
scale $ t_{rel} \approx (\alpha \, T \ln[1/\alpha])^{-1} $ arising from
infrared physics and hinges upon the separation of scales $ t_{rel} >>
\omega_p^{-1} $.  
\end{abstract} 
\section{Introduction}
The possibility of probing the quark-pluon plasma at the forthcoming
RHIC at BNL and LHC at CERN, has sparked an intense activity in the
understanding of collective excitations in ultrarelativistic plasmas
and their potential experimental signatures (for reviews
see\cite{qgp}-\cite{lebellac}). An important issue in this program is
a consistent assessment of the time scales for thermalization and
equilibration of the quasiparticles and  their distribution functions,
i.e. the damping and interaction
rates\cite{kapusta,lebellac,lifetime,htlnonabel,rate1}. At high
temperatures usual perturbation theory breaks down, but a consistent
resummation program developed by Braaten and
Pisarski\cite{rob2}-\cite{blaizot1} provides a systematically improved
perturbative expansion known as the hard thermal loop resummation
(HTL). The HTL resummation incorporates screening corrections in a
gauge invariant manner and  is sufficient 
to render finite the damping rate of excitations at rest in the
plasma\cite{zeromom} and transport cross sections\cite{transport}. 

However these corrections are not sufficient to cure the infrared
sensitivity of the damping rate of hard charged 
excitations which is dominated by the exchange of quasistatic magnetic
photons and or  gluons\cite{robinfra,rate8}. Whereas 
in QCD it is expected that a non-perturbative magnetic mass will
provide an infrared cutoff and ameliorate the 
infrared sensitivity of the damping rate for hard charged particles\cite{robinfra,rate8},
in QED the transverse photons do not acquire 
a magnetic mass, screening is only dynamical through Landau damping
and therefore the infrared singularities remain, possibly to all
orders. This infrared sensitivity in the 
case of QED has led some authors to question the validity of the
quasiparticle description and exponential relaxation that 
is the main concept behind the calculation of the damping
rate\cite{kobes,pilon}. 

These questions were recently clarified by the implementation of a
Bloch-Nordsieck (eikonal) resummation of the infrared divergent
diagrams in QED\cite{iancu,taka} which leads to an anomalous real-time
relaxation of the electron propagator of the 
form $ \approx \exp[-\alpha \, T\, t\ln(t\,\omega_p)] $. This result has been
recently confirmed  in scalar QED (SQED) by implementing a dynamical
renormalization 
group resummation directly in real time\cite{boyrgir}. The advantage
of this dynamical renormalization group approach 
is that it provides a  real-time description of relaxation bypassing
the limitation of the quasiparticle interpretation. 
In reference\cite{boyrgir} the equivalence of the dynamical
renormalization group to well studied situations at zero 
temperature was established: for cases with infrared threshold
singularities at zero temperature it is equivalent to the
Bloch-Nordsieck and the usual Euclidean renormalization group
resummation. In the absence of infrared threshold singularities it
leads to the real time evolution obtained by more conventional methods
(for a more complete comparison the reader is referred
to\cite{boyrgir}). For SQED at finite temperature it leads to the
relaxation of the single particle propagator which agrees exactly with
the results of the eikonal (Bloch-Nordsieck) approximation\cite{iancu}
in QED.  
This is one more example of the similarities between QED, QCD and
SQED\cite{rebhan}  in lowest order in the HTL resummation.  

In this article we explore a different but related question: since in
hot QED and in SQED  the infrared divergences associated with the
emission and absorption of quasi-static magnetic photons lead to
anomalous, non-exponential relaxation of the propagator of charged
excitations\cite{iancu,taka,boyrgir}, what is the kinetic equation
that describes the relaxation of the {\em distribution function} for
these excitations?. This kinetic equation would be the equivalent 
of the Boltzmann equation for the distribution function. 

When the quasiparticle picture is valid, the relaxation time
approximation leads to the  linearized equation  
$\delta \dot{N}_k(t) = -\gamma_k \; \delta N_k(t)$ with $\delta N_k(t)$
is the departure from equilibrium of the quasiparticle distribution
function and  the relaxation rate $\gamma_k$ (inverse of the
relaxation time) is time independent and simply related to the
damping rate for single quasiparticles $\Gamma_k$:  
$\gamma_k= 2\, \Gamma_k$\cite{kapusta,lebellac,lifetime}. If the single
quasiparticle propagator is {\em not} exponentially damped it is
reasonable to expect that the 
linearized kinetic equation for the distribution function will require
a time dependent generalization of $\gamma_k$.  

The goals of this article are a): a derivation of the proper kinetic
equation for the case of hard charged (quasi) particles, b)
interpretation of the relaxation time approximation and  comparison to
the relaxation of the single (quasi) particle Green's function. 

 We study these issues within the context of SQED to make contact with
 previous results\cite{boyrgir}. This model {\em is} relevant to study
 the physics of the QGP  because to lowest order in $\alpha$ SQED has
 the same infrared divergence and HTL structure as both QED and
 QCD\cite{rebhan}.  

{\bf The strategy:} Our strategy leading to the kinetic equation
follows closely the derivation of the Boltzmann 
equation presented in\cite{boykin,boyhtl}. It begins by defining a
suitable number operator $N_k$. In the case under consideration that
of a hard scalar with momentum $ k \geq T $ the collective mode and the
(renormalized) particle are indistinguishable\cite{rebhan,lebellac}, thus the
number operator is the usual operator associated with asymptotic
states. 
 The time derivative of this operator is
obtained from the Heisenberg equations of motion, and its average over
an initial density matrix is performed using 
 perturbation theory in terms of the non-equilibrium
propagators\cite{boykin,boyhtl}. We find that the time evolution 
of $ N_k $ obtained in this perturbative expansion contains logarithmic
{\em secular} terms in time. The dynamical renormalization 
group program is invoked to resum these secular terms. The resummed
distribution function obeys a dynamical renormalization group equation
which is recognized to be the generalization of the Boltzmann
equation\cite{nos}.  

{\bf The results:} There are two main results of our study, one of
general scope and the other particular to the 
anomalous kinetics of hard charged particles in SQED (but likely to be
shared by QED and QCD in lowest order in HTL resummation): a) we propose a microscopic
derivation of quantum kinetic equations beginning from a
non-equilibrium perturbative expansion in real time and using the
dynamical renormalization group to resum the secular terms. This
formulation allows to include other resummation schemes such as HTL
directly in the derivation of 
the kinetic equation. b) More specifically to the problem of the
kinetics of the distribution function for hard charged particles, we
have focused on SQED as a model that bears many of the relevant
features of QED and QCD. We find that the linearized version of the
kinetic equation (relaxation time approximation) for this distribution
function evolves in time 
as $ \delta n_k(t) = \delta n_k(t_o) \; e^{-2\,\alpha T\,
(t-t_o)\,\ln\bar{\mu}(t-t_o)} $ for $ t>>t_o $ (with $\alpha
= e^2/(4\pi) $ the `fine 
structure constant' and $\bar{\mu}\approx \omega_p$). This must be
compared to the real-time evolution of the hard scalar propagator
$G_k(t) \approx   
e^{-\alpha\, T\, t\,\ln\bar{\mu}t}$ \cite{boyrgir} which reveals a similar
relation between the relaxation time scales of the single 
particle Green's function and that of the distribution function as in
the usual quasiparticle picture but in a manner that is {\em not}
associated with pure exponential relaxation. 

A similar anomalous relaxation has been found by\cite{iancu} in spinor
QED via the Bloch-Nordsieck (eikonal) approximation.  

The article is organized as follows: in section II we introduce the
theory and describe the perturbative framework that leads to the
kinetic equation to ${\cal O}(\alpha)$. The different contributions
are interpreted as in-medium processes. 
In section III we focus on the relaxation time approximation
(linearized kinetic equation), recognize the secular terms 
arising in the perturbative expansion and implement the dynamical
renormalization group resummation of these secular 
terms in the asymptotic long time limit. In section IV we provide an
interpretation of the resummation scheme resulting from the dynamical
renormalization group and generalize the solution to include short
time transients. Section V summarizes our 
 conclusions, provides an assessment of the potential impact of these
results and poses other questions.

\section{The kinetic equation} 

We propose to study the relaxation of hard charged scalars in scalar
QED as a prelude to studying the more technically 
involved cases of QED and QCD. Scalar QED shares many of the important
features of QED and QCD in leading order in the HTL
resummation\cite{rebhan}. Furthermore, the infrared physics in QED
captured in the eikonal approximation (Bloch-Nordsieck) as clearly
explained in\cite{iancu} has been reproduced recently via the
dynamical renormalization group in SQED\cite{boyrgir}, thus lending
more support to the similarities of both theories at least in leading
HTL order.  

In the  Abelian theory under consideration, it is rather
straightforward to implement a  
gauge invariant formulation by projecting the Hilbert space on states
annihilated by Gauss' law. Gauge invariant operators can be constructed
and the Hamiltonian and Lagrangian can be written in terms of these. The 
resulting Lagrangian is exactly the same as that in Coulomb
gauge\cite{boyhtl} (for more details see\cite{gaugeinv}) 
 and is given by
\begin{eqnarray}  
{\cal L}=&&\partial_\mu\Phi^\dagger\,\partial^\mu\Phi -m^2 \Phi^{\dagger}\Phi  
+\frac{1}{2}\partial_\mu \vec{A}_T\cdot\partial^\mu\vec{A}_T 
-e\vec{A}_T\cdot\vec{j}_T 
-e^2\vec{A}_T\cdot\vec{A}_T\; \Phi^\dagger\Phi +\nonumber \\ 
&&+\frac{1}{2}\left(\nabla A_0 \right)^2+ {e^2}A^2_0 \; \Phi^{\dagger}\Phi+ 
eA_0\;\rho\; , \nonumber \\ 
\vec{j}_T=&&i(\Phi^\dagger\vec{\nabla}_T\Phi-\vec{\nabla}_T\Phi^\dagger\Phi)
\quad ; \quad \rho = -i\left(\Phi
\dot{\Phi}^{\dagger}-{\Phi}^{\dagger}\dot{\Phi}\right)\;.  \nonumber
\end{eqnarray} 
where we have traded the instantaneous Coulomb interaction for a gauge
invariant Lagrange 
multiplier field $ A_0 $ which should not be confused with a time component
of the gauge field. $ \vec{A}_T $ is the transverse component satisfying
$ \vec{\nabla}\cdot \vec{A}_T(\vec x,t)=0 $. 
Since we are only interested in obtaining the infrared behavior arising
from finite temperature effects we do not introduce the renormalization
counterterms to facilitate the study, although these can be systematically included in our formulation\cite{boyrgir}. The finite temperature behavior is
ultraviolet finite. The non-equilibrium generating
functional requires the fields on the forward and backward time
branches as discussed in \cite{boyhtl} (for more details on the non-equilibrium formulation see\cite{ctp}-\cite{tadpole1}). 

We anticipate that the Coulomb interaction will not be relevant to the
infrared physics because the longitudinal photon will be screened
with a screening mass $m_s \propto eT$. Only the transverse photons will
lead to infrared divergences\cite{robinfra,iancu,boyrgir} and
therefore we neglect the contribution from longitudinal photons (the
Coulomb interaction). Furthermore we will consider a neutral system
with vanishing chemical potential. It is convenient to perform a
spatial Fourier transform at some given time and write 
\begin{eqnarray}
\Phi(\vec x) & = & \int {d^3k\over (2\pi)^{3/2}}\;\phi(\vec
k)\; e^{i\vec k \cdot \vec x} 
\; \; ; \; \; \phi(\vec k)=
\frac{1}{\sqrt{2\omega_k}}\left[a_k + b^{\dagger}_k\right] \nonumber\\
\dot{\Phi}(\vec x) & = & \int {d^3k\over (2\pi)^{3/2}}\;
\dot{\phi}(\vec k)\; e^{i\vec k \cdot \vec x} \; \; ; \; \; 
\dot{ \phi}(\vec k) = -i\sqrt{\omega_k \over 2}
\left[a_k - b^{\dagger}_k\right] \nonumber 
\end{eqnarray} 
And the spatial Fourier transform for the transverse gauge field
$$
\vec A_T(\vec x) = \int {d^3k\over (2\pi)^{3/2}}\;
\vec{\cal A}_T(\vec k) \; e^{i\vec k \cdot \vec x}\;\; .
$$
Since we are interested in hard charged massive scalars, $\omega_k
\approx k$ the collective modes are indistinguishable from the single
particle excitations\cite{rebhan,lebellac}. The soft collective modes require a definition
of the quasiparticle number operator, for which the formulation
recently proposed in terms of quasiparticles\cite{welquasi} could
prove useful. This case will be explored elsewhere.  

The number of positively charged scalars (which at zero chemical
potential is equal to the number of negatively charged scalars) is
then given by the following Heisenberg operator 
\begin{eqnarray}
n_+(k,t) = a^{\dagger}_k(t) a_k(t) & = &  \frac{1}{2\omega_k}
\left[\dot{\phi}^{\dagger}(\vec k,t)\dot{\phi}(\vec k,t)+ \omega^2_k\;
\phi^{\dagger}(\vec k,t)\phi(k,t) \right. \nonumber \\ 
&+& \left. i\omega_k \left(\phi^{\dagger}(\vec k,t)\dot{\phi}(\vec
k,t)- \dot{\phi}^{\dagger}(\vec k,t) \phi(\vec k,t)\right)\right] \nonumber
\end{eqnarray}
Using the Heisenberg equations of motion we obtain
\begin{eqnarray}
\dot{n}_+(k,t) & = & \frac{e}{\omega_k } 
\int {d^3q\over (2\pi)^{3/2}}{\vec K}_T(\vec q)\cdot \vec{\cal
A}_T(\vec q,t) \phi^{\dagger}(\vec k + \vec
q,t)\left(\frac{\partial}{\partial t}- i\omega_k \right)\phi(\vec k,t)
 \nonumber \\ 
&+ &  \frac{e}{\omega_k }\left[\left(\frac{\partial}{\partial
t}+ i\omega_k\right)\phi^{\dagger}(\vec k,t)\right]
\int {d^3q\over (2\pi)^{3/2}}
 {\vec K}_T(\vec q)\cdot \vec{\cal A}_T(\vec q,t) \phi(\vec k-\vec q,t)
\nonumber\\ 
{\vec K}_T(q) & = & \vec k- \hat{\vec q}~(\vec k \cdot 
\hat{\vec q})\nonumber 
\end{eqnarray}  
\noindent We need to take the expectation value of this Heisenberg
operator in the initial density matrix. The following identities prove
useful  
\begin{eqnarray}
Tr\left[A(t)B(t)C(t)\rho(t_i)\right] & = &
Tr\left[B(t)C(t)\rho(t_i)A(t)\right] = \langle B^+(t)C^+(t)A^-(t)
\rangle \nonumber\\ 
& = & Tr\left[ C(t)\rho(t_i)A(t)B(t)\right] = \langle C^+(t) A^-(t)
B^-(t)\rangle \nonumber
\end{eqnarray}
where the $\pm $ correspond to operators defined on the forward ($+$) and
backward ($-$) time branches, which are obtained as functional derivatives
with respect to sources of the forward and backward time evolution
operators\cite{ctp}-\cite{tadpole1}. 
These identities allow to extract the time derivatives outside of the
expectation values and to avoid the potential 
Schwinger terms associated with  time ordering. Now the expectation
value of $\dot{n}_+(k,t)$ given above can be written as follows
\begin{eqnarray}
\langle \dot{n}_+(k,t) \rangle & = & \frac{e}{\omega_k } \left(
\frac{\partial}{\partial t'}- i\omega_k \right) 
\int {d^3q\over (2\pi)^{3/2}} K_{i,T}(q) 
\langle {\cal A}^-_{i,T}(\vec q,t) \phi^{\dagger, -}(\vec k+\vec q,t)
\phi^+(\vec k,t') \rangle\left|_{t=t'}\right. + \nonumber \\ 
& & \frac{e}{\omega_k} \left(
\frac{\partial}{\partial t'}+ i\omega_k \right) \int {d^3q\over
(2\pi)^{3/2}} K_{i,T}(q) 
\langle {\cal A}^+_{i,T}(\vec q,t) \phi^+(\vec k-\vec q,t) 
\phi^{\dagger,-}(\vec k,t') \rangle\left|_{t=t'} \right.\nonumber
\end{eqnarray} 
This expectation value is  computed in non-equilibrium perturbation
theory in terms of 
the non-equilibrium propagators and vertices. Since we are interested
in the infrared region in the internal loop momenta, we must consider the
HTL resummed photon propagator, but the scalar propagator need not be
resummed because $k \geq T$ and the scalar in the loop is hard and
massive. The required non-equilibrium propagators are the
following\cite{boyhtl,boyrgir,tadpole1} 
$\bullet\text{Scalar Propagators (zero chemical potential)}$ 
$$
{\langle}{\phi}^{(a)\dagger}(\vec{p},t){\phi}^{(b)}(-\vec{p}, 
t^{\prime}){\rangle}=-iG_p^{ab}(t,t^\prime)$$
where $(a,b)\;\in\{+,-\}$. 
\begin{eqnarray} 
&&G_p^{++}(t,t^\prime)=G_p^{>}(t,t^{\prime})\Theta(t-t^{\prime}) 
+G_p^{<}(t,t^{\prime})\Theta(t^{\prime}-t)\; , \nonumber
\\ 
&&G_p^{--}(t,t^\prime)= G_p^{>}(t,t^{\prime})\Theta(t^{\prime}-t)+ 
G_p^{<}(t,t^{\prime})\Theta(t-t^{\prime})\;, \nonumber
\\ 
&&G_p^{\pm\mp}(t,t^\prime)=G_p^{<(>)}(t,t^{\prime})\;, \nonumber \\ 
&&G_p^{>}(t,t^{\prime})=\frac{i}{2\omega_p}\left[ 
(1+n_p)\;e^{-i\omega_p(t-t^\prime)} 
+n_p\;e^{i\omega_p(t-t^\prime)}\right]\;,
\nonumber\\ 
&&G_p^{<}(t,t^{\prime})=\frac{i}{2\omega_p}\left[ 
n_p\;e^{-i\omega_p(t-t^\prime)} 
+(1+n_p)\;e^{i\omega_p(t-t^\prime)}\right]\;,
\nonumber\\ 
&&n_p= 
\frac{1}{e^{\beta\omega_p}-1}\;.  \nonumber
\end{eqnarray} 
$\bullet\text{HTL dressed Photon Propagators}$\cite{lebellac,boyrgir} 
$$
{\langle}{\cal A}^{(a)}_{Ti}(\vec{q},t){\cal A}^{(b)}_{Tj}(-\vec{q}, 
t^{\prime}){\rangle}=-i{\cal G}_{ij}^{ab} 
(\vec q ;t,t^\prime) 
$$
\begin{eqnarray} 
&&{\cal G}_{ij}^{++}(q;t,t^\prime)={\cal P}_{ij}(\vec{q}) \;
\left[{\cal G}_q{>}(t,t^{\prime})\Theta(t-t^{\prime}) 
+{\cal G}_q^{<}(t,t^{\prime})\Theta(t^{\prime}-t) \right]\;,\nonumber \\ 
&&{\cal G}_{ij}^{--}(q;t,t^\prime)= {\cal P}_{ij}(\vec{q}) \;
\left[{\cal G}_q^{>}(t,t^{\prime})\Theta(t^{\prime}-t) 
+{\cal G}_q^{<}(t,t^{\prime})\Theta(t-t^{\prime}) \right]\;, 
\nonumber\\
&&{\cal G}_{ij}^{\pm\mp}(q;t,t^\prime)={\cal P}_{ij}(\vec{q}) \;
{\cal G}_q^{<(>)}(t,t^{\prime})\;, \nonumber\\ 
&&{\cal G}_q^{>}(t,t^{\prime})=\frac{i}{2}\int dq_o
\;\tilde{\rho}_T(q_o;\vec q)\left[ (1+N_{q_o})e^{-iq_o(t-t^\prime)}  
+N_{q_o} e^{iq_o(t-t^\prime)}\right]\;,\nonumber
\\ 
&&{\cal G}_q^{<}(t,t^{\prime})=\frac{i}{2}\int dq_o\;
\tilde{\rho}_T(q_o;\vec q)\left[  
N_{q_o}\;e^{-iq_o(t-t^\prime)} 
+(1+N_{q_o})\;e^{iq_o(t-t^\prime)}\right],\nonumber\\
&&N_{q_o}=\frac{1}{e^{\beta q_o}-1}\;. \nonumber
\end{eqnarray} 
Where we have used the properties\cite{boyrgir,lebellac}
\begin{equation}
\tilde{\rho}_T(-q_o, q) = - \tilde{\rho}_T(q_o,q) ~~; ~~ N(-q_o)=
-[1+N(q_o)] \label{properties}  
\end{equation}
\noindent The HTL spectral density is given by\cite{boyrgir,rebhan,boyhtl}
\begin{eqnarray}
\tilde{\rho}_T(q_0,q) & = &
\frac{1}{\pi}\frac{\Sigma_I(q_0,q)\;\Theta(q^2-q^2_0)}{\left[q^2_0-q^2
-\Sigma_R(q_0,q)\right]^2+\Sigma^2_I(q_0,q)} +\mbox{sign}(q_0)\;
Z(q)\;\delta(q^2_0-\omega^2_p(q)) \label{rhophoton} \\ 
\Sigma_I(q_0,q) & = & \frac{\pi e^2
T^2}{12}\frac{q_0}{q}\left(1-\frac{q^2_0}{q^2}\right) \label{landaucut} \\ 
\Sigma_R(q_0,q) & = & \frac{ e^2
T^2}{12}\left[2\;\frac{q^2_0}{q^2}+\frac{q_0}{q}
\left(1-\frac{q^2_0}{q^2}\right)\ln\left|\frac{q_0+q}{q_0-q}\right|\right]     
\nonumber
\end{eqnarray}
where $\omega_p(q)$ is the plasmon pole and $Z(q)$ its (momentum dependent)
residue, which will not be relevant for the following discussion. The 
important feature of this HTL resummed spectral density is its support below
the light cone, i.e. for $q^2> q^2_0$, the imaginary part
(\ref{landaucut}) originates in the process of Landau
damping\cite{robinfra} from scattering of quanta in the medium.  
Above ${\cal P}_{ij}(\vec{q})$ is the transverse projection operator: 
$$
{\cal P}_{ij}(\vec{q})=\delta_{ij}-\frac{q_i q_j}{q^2}\;.
$$
Using the above expressions for the non-equilibrium propagators, and after
some tedious but straightforward algebra, we find the expectation value
$\langle \dot{n}_k(t) \rangle$ to lowest order in perturbation theory
${\cal O}(e^2)$ is given by 
\begin{eqnarray}
\langle \dot{n}_k(t) \rangle & = & \frac{e^2 k^2}{4\pi^2 \omega_k}
\int_0^{\infty} \frac{q^2 dq}{\omega_{\vec k+\vec q}}
\int^1_{-1}d\cos\theta\;(1-\cos^2\theta)\int_{-\infty}^{\infty}
dq_o~\tilde{\rho}_T(q_o,q) 
\int^t_{t_i}dt' \times \nonumber \\
&&\left\{
\left[ (1+N_{q_o})(1+n_{\vec k+\vec q})(1+n_k) - N_{q_o}\; n_{\vec k+\vec
q}\;n_k\right] \cos(q_o+\omega_{\vec k+\vec
q}+\omega_k)(t-t')+\right. \nonumber \\ 
&&\left. \left[ N_{q_o}(1+n_{\vec k+\vec q})(1+n_k) -
(1+N_{q_o})\;n_{\vec k+\vec q}\;n_k\right] \cos(-q_o+\omega_{\vec
k+\vec q}+\omega_k)(t-t')+ \right. \nonumber \\ 
&&\left. 2\left[N_{q_o}\;n_{\vec k+\vec q}\;(1+n_k) -
(1+N_{q_o})(1+n_{\vec k+\vec q})\;n_k\right] \cos(q_o+\omega_{\vec
k+\vec q}-\omega_k)(t-t')\right \}\label{finalndot} 
\end{eqnarray} 
where  $\theta$ is the angle
between $\vec k$ and $\vec q$. The different contributions have a very
natural interpretation in terms of $\mbox{gain}- \mbox{loss}$
processes. The first term in  
brackets corresponds to the process $0 \rightarrow \gamma^* + s + \bar{s}$ 
minus the process $\gamma^* + s + \bar{s} \rightarrow 0$, the second term 
corresponds to $\gamma^* \rightarrow s+ \bar{s}$ minus $s+ \bar{s}
\rightarrow \gamma^*$, and the last term corresponds to the scattering
in the medium  
$\gamma^* + s \rightarrow s$ minus the inverse process $s \rightarrow
\gamma^* + s$ where $\gamma^*$ refers to the HTL- dressed photon and
$s, \, \bar{s}$ refer to the charged quanta of the scalar field
$\Phi$.   

\section{Relaxation time approximation: secular terms}

We will assume that there is an equilibrium solution for the
distribution function and that at time $t=t_0$ the distribution
function for a fixed mode
$k$ is disturbed slightly off equilibrium so that $n_k(t<t_0)=n_k^{eq};~
n_k(t=t_0)=n_k^{eq}+\delta n_k(t_0)$ while the rest of the modes
remain in equilibrium. We want to study the time evolution  of the
perturbed distribution in the linearized approximation. 
This leads to a kinetic
equation (\ref{finalndot})  that is linear in $\delta n_k$, since consistent with
the condition that only the mode of wavevector $k$ is off equilibrium,  we set $ \delta n_{\vec k+\vec q} = 0 $ for $ {\vec q} \neq 0 $ and the 
point $ {\vec q} = 0 $ does not contribute because the integration measure  vanishes there. 

This is known as the relaxation time
approximation. Only in this approximation the relaxation of the
distribution function for the number of particles is related to the
relaxation of single particle Green's function.  Since the propagators
entering in 
the perturbative expansion of the kinetic equation are in terms of the
distribution functions at the initial time the time integration can be
done straightforwardly leading to the following linearized equation
\begin{eqnarray}
\delta{\dot n}_k(t) & = & \delta n_k(t_0)\; \frac{e^2 k^2}{4\pi^2
\omega_k} \int_0^{\infty} \frac{q^2 dq}{\omega_{\vec k+\vec q}}
\int^{+1}_{-1}(1-\cos^2\theta)\;d\cos\theta\;\int_{-\infty}^{+\infty}
dq_o \;\tilde{\rho}_T(q_o,q) 
\times \nonumber \\
&& \left\{2
\left[ 1+ N_{q_o}+n_{\vec k+ \vec q}\right]
\frac{\sin(q_o+\omega_{\vec k + \vec
q}+\omega_k)(t-t_0)}{q_o+\omega_{\vec k + \vec 
q}+\omega_k}\right. \nonumber \\ 
&-&\left. 2\left[ 1+ N_{q_o}+n_{\vec k+ \vec q}\right]
\frac{\sin(q_o+\omega_{\vec k + \vec
q}-\omega_k)(t-t_0)}{q_o+\omega_{\vec k + \vec q}-\omega_k}\right\} 
\label{relaxtimeappx}
\end{eqnarray}
\noindent where we have used the properties (\ref{properties}) to
combine the first and second terms in the kinetic equation 
(\ref{finalndot}).   

Before proceeding  further, it is illustrative to analyze the above
equation in the long time limit $t>>t_o$.  
{\em If} the denominators in (\ref{relaxtimeappx}) have isolated
zeroes in the region of integration we can use the 
approximation ${\sin[\Omega\tau]}/{\Omega} \buildrel{\tau \rightarrow
\infty}\over\approx \pi \delta(\Omega)$   
which is the usual approximation leading to Fermi's Golden rule. Thus
{\em if} there are no singularities arising from the integrals and if
the resulting delta functions are satisfied  
one finds a linear secular term in time upon integrating in time the
rate equation (\ref{relaxtimeappx}). This is a perturbative signal of 
pure exponential relaxation at long times\cite{nos}. However in the  case
under consideration there are  infrared singularities in the
integrand\cite{iancu,boyrgir} and the long-time limit must be studied
carefully.  

For this purpose it is convenient to introduce the following spectral
density
\begin{eqnarray}
\rho(k;\omega) & = &  -\frac{e^2k^2}{\pi^2} \int_0^{\infty}
\frac{q^2dq}{\omega_{\vec k+\vec q}}\int_{-1}^1(1-\cos^2\theta)\;
d\cos\theta \; \int_{-\infty}^{\infty} dq_o\;
\tilde{\rho}_T(q_o,q)\times \nonumber \\ 
& & \left[1+N(q_o)+n_{\vec k+\vec
q}\right]\;\delta(\omega-q_o-\omega_{\vec k+\vec q})\label{specfunc} 
\end{eqnarray}
which is  the same as that studied within the
context of the relaxation of the amplitude of a mean field in\cite{boyrgir}
and in the eikonal approximation\cite{iancu}.

In terms of this spectral density we obtain the time derivative of
distribution function in the form
\begin{eqnarray}
\delta{\dot n}_k(t) & = & -\alpha\; \Gamma_k(t) \;\delta n_k(t_0) \nonumber \\
\alpha \Gamma_k(t) & = & -\frac{1}{2\omega_k}\int d\omega\;
\rho(k;\omega)
\left[\frac{\sin[(\omega-\omega_k)(t-t_o)]}{(\omega-\omega_k)}+ (\omega
\rightarrow -\omega) \right] ~~;~~ \alpha= \frac{e^2}{4\pi}\label{gammadef}
\end{eqnarray} 
which upon integrating in time with initial condition at $t_o$ leads to
the form
$$
\delta n_k(t)  =  \delta n_k(t_0)\left\{1- \alpha
\int^t_{t_o}\Gamma_k(t')\;dt'\right\} 
$$
We focus on the case of hard external momemtum $k \approx T$ but 
on  the infrared region of the loop integral $q_o,~q \leq eT$ in the
spectral function (\ref{specfunc})\cite{iancu,boyrgir}. This is the
region dominated by the exchange of soft, HTL resummed transverse
photons\cite{robinfra,iancu} and that dominates the long time
evolution of the distribution function. 
 In this region we can replace $\omega_k \approx k  \; ;  \;
\omega_{\vec k + \vec q} \approx k+q \, \cos\theta $.  

Potential secular terms (growing in time) could arise in the long time
limit $ t>>t_o $  whenever the denominators in
(\ref{gammadef}) vanish, i.e. for  the region of frequencies $\omega
\approx \omega_k \approx k$ and $\omega \approx -\omega_k \approx
-k$. For 
$\omega \approx \omega_k \approx k$ we see that the argument of the delta
function in (\ref{specfunc}) vanishes in the region of the Landau
damping cut of the exchanged transverse photon $q^2_o < q^2$ and
contributes to the infrared behavior. On the other hand, for $\omega
\approx -\omega_k \approx -k$ the delta function 
in (\ref{specfunc}) is satisfied for $q_o \approx -2k$, and this
region gives a negligible contribution to the long time
dynamics. Therefore only the first term in (\ref{gammadef}) (with
$\omega-\omega_k$) contributes in the long time limit. This term is
dominated by the Landau damping region of the spectral density of the
exchanged soft photon given by (\ref{rhophoton}) since for $\omega \approx \omega_k$ the argument of the delta function is $q_o+q~\cos(\theta)$ and this is the region where the imaginary part (\ref{landaucut}) has support. The second contribution
(with $\omega + \omega_k$) oscillates in time and is always bound and
perturbatively small.   

In the hard limit $\omega_k \approx k$, and for $q_o; q \leq eT<<T$ we neglect the contribution from $n_{\vec k+\vec q}$ (hard and massive scalar) and replace
$N(q_o) \approx T/q_o$. The spectral density is found to be given
by\cite{boyrgir} 
$$
\rho(k;\omega) \buildrel{\omega \to k}\over= \frac{e^2 kT}{\pi^2}
\ln\left|\frac{\omega-k}{\mu}\right|+ {\cal O}(\omega-k)
$$
where $\mu$ is an infrared cutoff $\mu \approx \omega_p \approx
eT$\cite{robinfra,iancu,boyrgir}. 

In the limit $ \mu (t-t_o) >>1 $ we  find\cite{boyrgir}
\begin{equation}
\int^t_{t_o}\Gamma_k(t')dt' \buildrel{\mu (t-t_o) >>1}\over= 2 
T\, (t-t_o) \ln[\bar{\mu}(t-t_o)]+\mbox{non-secular~terms}\label{secular} 
\end{equation} 
with $ \bar{\mu}= \mu \exp[\gamma_E -1] $ and $\gamma_E$ is the
Euler-Mascheroni constant.  
In lowest order in perturbation theory, the occupation number that enters
in the loops are those at the initial time.
Obviously perturbation theory breaks down at time scales $ t-t_o 
\approx \left|\alpha T\ln\alpha \right|^{-1}$ where we used $\bar{\mu}
\approx \omega_p \approx eT$. This situation is similar to the case of
a weakly damped harmonic oscillator when a perturbative solution in
terms of the damping coefficient is 
sought. Such a perturbative solution has secular terms that grow
linearly in time in lowest order, which reflect the perturbative
expansion of the damping exponential, i.e. $e^{-\gamma t} \approx 1-
\gamma t +\cdots$. This perturbative expansion breaks down at time
scales ${\cal O}(1/\gamma)$. The dynamical renormalization group is a
systematic 
generalization of multi-time scale analysis and sums the secular
terms, thus improving the perturbative expansion\cite{goldenfeld}. For
a discussion of the dynamical renormalization group 
in other contexts, including applications to quantum field theory
problems see\cite{goldenfeld}-\cite{qm}. 

 We now implement the dynamical renormalization group to sum the
 secular divergences and to improve the perturbative expansion using
 the formulation advanced in\cite{boyrgir}.  

To achieve this purpose we introduce a renormalization constant for
the occupation number that absorbs the secular divergences at a fixed
time scale $\tau$ and write 
\begin{equation}
\delta n_k(t_o) = \delta n_k(\tau) \; {\cal Z}(\tau,t_o) \; \; ; \; \; 
{\cal Z}(\tau,t_o)= 1 + \alpha \; z_1(\tau,t_o) + \cdots \label{renor}
\end{equation} 
and request that the coefficients $z_n$ cancel the secular divergences
proportional to $\alpha^n$ at a given time scale $\tau$. To lowest
order the choice
\begin{equation}
z_1(\tau,t_o) = \int^{\tau}_{t_o}\Gamma_k(t')\;dt' \label{zeta1}
\end{equation}
\noindent leads to the renormalized distribution function at time $t$ in terms of the updated occupation number at the time scale $\tau$
$$
\delta n_k(t)  =  \delta n_k(\tau)\left\{1- \alpha
\int^t_{\tau}\Gamma_k(t')\;dt'\right\} 
$$
However, the occupation number $\delta n_k(t)$ cannot depend on the
arbitrary renormalization scale $\tau$, this independence on the
renormalization scale leads to the renormalization group equation to
lowest order: 
\begin{equation}
\frac{\partial \delta n_k(\tau)}{\partial \tau} + \alpha\;
\Gamma_k(\tau)\; \delta n_k(\tau) =0  \; . \label{RGeqn} 
\end{equation}
This renormalization group equation is now clearly of the form of a
Boltzmann equation in the relaxation time approximation with a time
dependent rate. 

Now choosing the renormalization scale to coincide with the time $t$ in
the solution of (\ref{RGeqn}) 
as is usually done in the scaling analysis of the solutions to the
renormalization group equations,  we find that the distribution
function in the linearized approximation evolves in time in the 
following manner
\begin{equation}
\delta n_k(t)= \delta n_k(t_o)\; e^{-\alpha \int^t_{t_o}\Gamma_k(t')\;dt'}
\label{expsol}
\end{equation}
with the initial conditions 
\begin{equation}
\delta n_k(t=t_o) = \delta n_k(t_o) \; ; \; \; 
\delta{\dot n}_k(t)\left|_{t=t_o}\right. = 0  \label{inicond}
\end{equation}
in agreement with the perturbative expression (\ref{relaxtimeappx}).
In the long time limit $\overline{\mu}(t-t_o) >>1$ and using (\ref{secular}) we
find that the distribution function  relaxes towards equilibrium
as 
\begin{equation}
\delta n_k(t) \buildrel{\mu (t-t_o) >>1}\over=\delta n_k(t_o)
e^{-2\alpha T (t-t_o)\ln[(t-t_o)\bar{\mu}]} \label{linearrg}
\end{equation}
The exponent in (\ref{linearrg}) must be compared to that found for the
relaxation of the mean field $\langle \Phi(\vec k,t) \rangle$ in\cite{boyrgir}.  
We thus find that the distribution function approaches equilibrium as the
{\em square} of the mean field, a result which is in agreement with the
usual relation between the damping and the interaction rate. 

Furthermore,  the dynamical renormalization group resummation of the
perturbative expansion reveals a relaxation time scale for the
distribution function of the hard charged particles given by 
$$
t_{rel} \approx \left|\alpha T\ln\alpha\right|^{-1}\; .
$$

\section{Interpretation of the RG in kinetics}

The interpretation of eq.(\ref{secular}) and the renormalization
of the distribution function given by eq.(\ref{renor}) are clear:
having prepared the initial distribution at a time $t_o$ we evolve the
distribution forward in time using perturbation theory but during a
time scale $\tau$ such that the perturbative expansion is still valid,
i.e. $t_{rel}>>(\tau-t_o)$. Secular terms begin to dominate the perturbative
expansion at a time scale  $(\tau-t_o)>>\omega^{-1}_p$, if there is
a separation of time scales such that $t_{rel}>>(\tau-t_o)>>\omega^{-1}_p$
perturbation theory is reliable in this regime but secular terms appear
and can be isolated. Thus the perturbative expansion is valid for a large
time scale, at a given {\em renormalization} scale $\tau$ within this
interval the result of the perturbative expansion is to `reset' the
occupation number, from $\delta n_k(t_0)$ to $\delta n_k(\tau)$. Having
`reset' the occupation number at this time scale in which perturbation
theory is reliable, this new occupation number is used as an initial
condition to iterate forward in time using perturbation theory to
another time that again is within the perturbatively reliable region.  

 The renormalization
condition (\ref{renor}) 
is precisely the resetting of the occupation number. Now the
perturbative expansion has been improved and can be extended to longer
time scales by using the updated occupation number as an initial condition
at the scale $\tau$. The dynamical renormalization group equation (\ref{RGeqn})  is the differential form of the procedure of `resetting' followed by perturbative evolution in real time. This renormalization group equation (\ref{RGeqn}) is  identified with the Boltzmann equation in the
relaxation time approximation\cite{nos}.

We note that there is certain freedom in the choice of the `renormalization
counterterms' $z_n$ (\ref{renor}): rather than choosing $z_1$ as in
eq.(\ref{zeta1}) we could have chosen to cancel {\em only the secular 
part} given by (\ref{secular}) rather than the whole integral. In the
language of renormalization this choice would correspond to choosing
the counterterms to cancel only the {\em divergent} contribution, i.e.
we could have chosen

\begin{equation}
z_1 = 2 T(\tau-t_o) \ln[\bar{\mu}(\tau - t_o)] \label{zeta1sec}
\end{equation}

\noindent which would lead to the renormalization group equation
\begin{equation}
\frac{\partial \delta n_k(\tau)}{\partial \tau} + 
2\alpha T \,\left\{\ln[\bar{\mu}(\tau-t_o)]+1\right\}\;  \delta
n_k(\tau)=0 \label{rgeqsec}  
\end{equation} 
 
Now choosing the renormalization scale $\tau$ with the time $t$ in the
solution of (\ref{rgeqsec}) we would find the asymptotic form
$$
\delta n_k(t) = \delta n_k(t_o)\; e^{-2\alpha T (t
-t_o)\ln[(t-t_o)\bar{\mu}]} 
$$
Obviously the two choices (\ref{zeta1}) and (\ref{zeta1sec}) lead to the
same asymptotic behavior, because they only differ by non-secular
terms which are subleading in the asymptotic regime. The 
choice (\ref{zeta1}) also contains non-secular contributions which are 
bound in time and therefore always perturbative, i.e. {\em finite}
terms in the usual renormalization program.   The advantage of
choosing (\ref{zeta1}) is that the solution (\ref{expsol}) includes
transient effects and obeys the  
initial conditions (\ref{inicond}) explicitly. 

In this particular case, the validity of the renormalization group
approach hinges upon a separation of time scales: the time scale at
which the occupation number is reset, i.e. $\tau$ has to be close to  
the scale $t$ so that the perturbative expansion has been improved but
not too long so that the occupation number has 
changed much during this scale. Since the scale of the kernel is given
by the plasma frequency $\omega_p \approx eT$ which 
enters in the self-energy of the HTL resummed photon and the infrared
cutoff $\mu$, the separation of scales is guaranteed  in weak coupling
since $t_{rel} \approx \left|\alpha T \ln(\alpha)\right|^{-1} >>
\omega_p^{-1} \approx (eT)^{-1}$. This is a general statement for the
validity of {\em any} kinetic approach, that is, there must be a clear
separation of scales between the relaxation and the microscopic time
scales. 

This interpretation of kinetics and the Boltzmann equation via the dynamical renormalization group provides a natural answer to the question: What
does the Boltzmann equation sum??. In a typical kinetic approach one
writes a Boltzmann equation by computing in perturbation theory some transition matrix element which enters in an integral kernel. Obviously
the full solution of the Boltzmann equation is {\em non-perturbative} even
when the transition matrix element has been computed in perturbation
theory. In particular in the relaxation time approximation the coupling
constant appears in the solution in an exponential form via the relaxation time or damping rate. What our analysis in terms of the renormalization group reveals is that what is being summed by the Boltzmann equations are {\em secular terms in time}. 

In this article we have focused on the derivation of the kinetic
equation and the renormalization group resummation in the relaxation
time approximation to compare the relaxation of the distribution function
of hard charged scalars to that of the single particle Green's function\cite{iancu,boyrgir}. We relegate the treatment
of the full non-linear kinetic equations via the dynamical renormalization
in its full generality to a forthcoming article\cite{nos}.

\section{Conclusions, implications and more questions:}

Our goal with this article was to understand the relaxation of the distribution function of charged fields with hard momenta in a hot
gauge theory. The detailed study of the relaxation of the single
charged (quasi)particle Green's function in the eikonal (Bloch-Nordsieck) approximation in references\cite{iancu,taka} revealed an anomalous relaxation as a consequence of the infrared divergences associated with the emission and absorption of HTL resummed transverse photons. This novel form of relaxation of the single particle Green's function was confirmed
in terms of a dynamical renormalization group
resummation in\cite{boyrgir}.  

There are several important implications of these results:

\begin{itemize}

\item{Although in QED there is no magnetic mass, the analisis
of\cite{iancu,taka,boyrgir} reveals that the time scale for relaxation
of fast moving single particle charged excitations is given by 
$2|\alpha T \ln(\alpha)|^{-1}$. This is the {\em same} time scale that
arises in QCD after resummation and assuming that a magnetic mass
$m_{mag}<<m_D \approx gT$ cuts off the infrared divergences in the
damping  rate\cite{robinfra,rate8}. Thus the conjecture in\cite{robinfra} is
confirmed through a non-exponential relaxation. The advantage of a
real time analisis is that the time variable serves as an infrared cutoff.
Furthermore the renormalization group approach is not restricted to the
quasiparticle picture.}

\item{The  {\em distribution function} of fast charged
excitations in the linearized approximation relaxes also with an 
an anomalous, non-exponential form that asymptotically is the {\em square}
of the single particle Green's function. Therefore this provides another
confirmation of the relation between the `damping' and `interaction'
rates but with  non-exponential relaxation. The relaxation time scale
for the distribution function is given by 
$t_{rel} \approx |\alpha T \ln(\alpha)|^{-1}$, i.e. one half of the
relaxation time scale for the single particle Green's function. }  

\item{This result suggests that at very large energy when the QCD coupling
is small, i.e. in the initial stages of an Ultrarelativistic Heavy Ion
Collision, the thermalization rate of quarks could be as fast as that
of gluons. The in-medium effects that lead to the logarithmic enhancement
of the thermalization scale could compensate for the group factors that
result in a larger thermalization rate for gluons}.

\end{itemize}
 
There are still unresolved and important questions that deserve attention.
In particular the treatment of collective modes. Here we have focused on
hard excitations, and in the limit of large momenta $k>>eT$ collective and
single particle excitations are indistinguishable. Hence the definition of
the number of charged excitations coincides with that for asymptotic single
particle states. However for $k \leq eT$ we must treat the distribution
function of the collective modes in terms of an interpolating operator
that counts these collective modes. Furthermore for soft collective modes,
not only the internal photon but the charged particle propagator and the
vertices must be HTL resummed\cite{robinfra,iancu} and this situation
must be studied in detail. We hope to report on progress on some of
these issues along with a treatment of non-linear kinetics and resolution of pinch singularities with the dynamical
renormalization group in a forthcoming article\cite{nos}.

\section{acknowledgements}
The authors would like to thank R. Pisarski, J.-P. Blaizot, E. Iancu, E. Mottola and L. Yaffe for enlightening conversations and  comments. 
D. B. thanks the N.S.F. for partial support through grant PHY-9605186 and
LPTHE at the Universit\'e Pierre et Marie Curie and Denis Diderot for
hospitality.  The authors acknowledge support from NATO.


\end{document}